\documentclass[11pt,twoside]{article}

\usepackage[utf8]{inputenc}
\usepackage{amsmath, amssymb, amsfonts, amsthm}
\usepackage{graphicx}
\usepackage{float}
\usepackage{booktabs}
\usepackage{multirow}
\usepackage{arydshln}

\usepackage[colorlinks=true, urlcolor=blue, citecolor=black, linkcolor=black]{hyperref}

\usepackage{tikz, pgfplots}
\pgfplotsset{compat=1.18}
\usepgfplotslibrary{groupplots, fillbetween, external}
\usetikzlibrary{patterns, arrows.meta, positioning, shapes}
\usepackage{pgfplotstable}
\usepackage{subcaption}

\usepackage{enumitem}

\usepackage{bibunits}
\usepackage[authoryear]{natbib}
\defaultbibliographystyle{plainnat} 

\usepackage{lmodern}
\usepackage{microtype}
\usepackage{titlesec}
\titleformat{\section}{\normalfont\Large\bfseries}{\thesection}{1em}{}
\titleformat{\subsection}{\normalfont\large\bfseries}{\thesubsection}{1em}{}

\usepackage[font=small,labelfont=bf]{caption}
\usepackage{comment}
\usepackage{authblk}  

\usepackage{geometry}
\title{Clustering and surface distributions of buoyant particles in open-channel flows}

\author[1]{Ana Todorova}
\author[2,1]{Robert K. Niven}
\author[1]{Matthias Kramer\thanks{Corresponding author: \texttt{m.kramer@unsw.edu.au}}}

\affil[1]{School of Engineering and Technology (SET), The University of New South Wales, Canberra, Australia}

\affil[2]{Department of Mechanical Engineering, Auckland University of Technology, Auckland, New Zealand}

\date{January 2026}

\usepackage{geometry}
\usepackage{fancyhdr}
\usepackage[T1]{fontenc}
\usepackage{newtxtext,newtxmath} 

\usepackage{amsmath, amssymb}

\usepackage{microtype}
\titleformat{\section}
  {\normalfont\normalsize\bfseries}
  {\thesection}{1em}{}

\titleformat{\subsection}
  {\normalfont\normalsize\itshape}
  {\thesubsection}{1em}{}

\titleformat{\subsubsection}
  {\normalfont\normalsize\itshape}
  {\thesubsubsection}{1em}{}

\titlespacing*{\section}{0pt}{12pt plus 2pt minus 2pt}{4pt}
\titlespacing*{\subsection}{0pt}{10pt plus 2pt minus 2pt}{4pt}
\titlespacing*{\subsubsection}{0pt}{6pt plus 1pt minus 1pt}{2pt}

\usepgfplotslibrary{colormaps} 

\pgfplotsset{
    colormap={mygray}{
        rgb255(0cm)       = (139,0,0)      
        rgb255(0.167cm)   = (200,30,30)    
        rgb255(0.333cm)   = (255,100,100)  
        rgb255(0.5cm)     = (255,180,180)  
        rgb255(0.667cm)   = (255,255,255)  
        rgb255(0.75cm)    = (180,200,255)  
        rgb255(0.833cm)   = (100,150,255)  
        rgb255(0.917cm)   = (50,80,255)    
        rgb255(1.0cm)     = (0,0,139)      
    }
}

\usepackage{caption}

\begin{document}

\maketitle

\begin{abstract}

This study investigates the clustering behaviour and surface distributions of buoyant particles at the air–water interface in open-channel turbulent flow, focusing on the interplay between capillary attraction, hydrodynamic drag, and flow-driven lateral transport. Using controlled laboratory flume experiments, we systematically examine clustering dynamics for two particle types differing in size and density. To interpret the observed behaviour, we extend capillary-based clustering frameworks to open-channel flows by introducing a dimensionless clustering Weber number ($We_\mathrm{cl}$) that captures the balance between the flow-induced disruptive force and capillary attraction, providing a compact description of the observed clustering behaviour. In addition, we demonstrate that secondary currents play a central role in surface particle transport, producing systematic lateral accumulation that depends on channel aspect ratio. Together, these findings extend capillary-driven clustering theory to open-channel turbulence and reveal secondary currents as a key mechanism controlling particle surface distributions.

\end{abstract}


\section{Introduction}

Floating particles commonly accumulate at fluid interfaces, where their motion reflects a combination of interfacial physics and flow-driven transport. Such accumulations are relevant in contexts ranging from natural foams and algal aggregates to industrial emulsions and the transport of floating plastic debris in rivers and estuaries. 

Historically, clustering at interfaces has been investigated primarily in the absence of flow, often using colloidal particles or droplets \citep{nicolson1949interaction, chan1981interaction,botto2012capillary, kralchevsky2009colloid}. In these systems, particles deform the interface and generate overlapping menisci, which produce lateral capillary attraction, a phenomenon often referred to as the `Cheerios effect' \citep{vella2005cheerios}. A recent study by \citet{Shin2024DenseInterface} examined the interplay between disruptive force and interfacial capillary attraction using particles confined at the interface of quasi-two-dimensional turbulent layers. They showed that clustering behaviour can be organised using two dimensionless parameters; the areal fraction $\phi$, defined as the projected particle area normalised by the observation area, and a capillary number, defined as the ratio of flow-induced disruptive force to inter-particle capillary attraction. 

\begin{figure}[h!]
    \centering
    \begin{tikzpicture}
\begin{groupplot}[group style = {group size = 2 by 2,horizontal sep=1cm,vertical sep=0cm}]

\nextgroupplot[
    height = 6 cm,
    width = 7 cm,
    xlabel={\(We_\mathrm{cl}\)},
    ylabel={\(\phi\)},
    grid=major,
    xmode=log,
    log basis x=10,
    xmin=0.01, xmax=100,
    ymin=0, ymax=1,
    legend style={
        cells={align=left},
        anchor=north west,
        at={(1.15,1)},
        font=\small,
        align=left,
        legend cell align=left,
        legend columns=1
    }
]

\addlegendimage{area legend, fill=black!20, draw=none, opacity=0.9}
\addlegendentry{Capillary raft}

\addlegendimage{area legend, fill=gray!10, draw=none, opacity=0.9}
\addlegendentry{Capillary cluster}

\addlegendimage{area legend, fill=teal!50!black!50, draw=none, opacity=0.9}
\addlegendentry{Hydrodynamic dispersion}

\addlegendimage{area legend, fill=orange!20!white, draw=none, opacity=0.9}
\addlegendentry{Hydrodynamic breakup}

\addlegendimage{dashed,color=black,
    line width=1.5}
\addlegendentry{\shortstack{Theoretical packing \\limit (2D circles)}}


\addlegendimage{ mark=*, 
    color=brown,
    draw=black,
    only marks,
    mark size=2pt}
\addlegendentry{{Present study}}



\addplot [
    draw=none,
    fill=gray!10,
    opacity=0.8
] coordinates {
    (0.01, 0)
    (1, 0)
    (1, 0.4)
    (0.01, 0.4)
    (0.01, 0)
};

\addplot [
    draw=none,
    fill=orange!20!white,
    opacity=0.9
] coordinates {
    (1, 0)
    (1000, 0)
    (1000, 0.4)
    (1, 0.4)
    (1, 0)
};

\addplot [
    draw=none,
    fill=black!20,
    opacity=0.9
] coordinates {
    (0.01, 0.4)
    (1, 0.4)
    (1, 0.9069)
    (0.01, 0.9069)
    (0.01, 0.4)
};

\addplot [
    draw=none,
    fill=teal!50!black!50,
    opacity=0.9
] coordinates {
    (1, 0.4)
    (1000, 0.4)
    (1000, 0.9069)
    (1, 0.9069)
    (1, 0.4)
};

\addplot[color=black,
    line width=1.5,
    solid
  ] coordinates {
    (1,0)
    (1,1)
  };

\addplot[color=black,
    line width=1.5,
    solid
  ] coordinates {
    (0.01,0.4)
    (1000,0.4)
  };

  \addplot[dashed,color=black,
    line width=1.5] coordinates {
    (0.01,0.9069)
    (1000,0.9069)
  };

\addplot[
    only marks,
    mark=*,
    mark size=2.5pt,
    color=brown,
    mark options={draw=black, line width=0.6pt}, 
    clip mode=individual
]
table[row sep=\\]{
x y \\
0.018844 0.07\\
0.085666 0.07\\
0.162654 0.07\\
0.251972 0.07\\
0.377732 0.07\\
0.903222 0.07\\
1.149298 0.07\\
1.380636 0.07\\
};

\addplot[
    only marks,
    mark=*,
    mark size=2.5pt,
    color=brown,
    mark options={draw=black, line width=0.6pt}, 
    clip mode=individual
]
table[row sep=\\]{
x y \\
0.005969 0.107\\
0.044208 0.107\\
0.082962 0.107\\
0.137086 0.107\\
0.204829 0.107\\
};

\node[fill=gray!10,fill opacity=0 ,text opacity=0.35] at (axis cs:0.1,0.2) {\small\shortstack{Capillary\\cluster}};

\node[fill=orange!20,fill opacity=0,text opacity=0.35] at (axis cs:10,0.2) {\small\shortstack{Hydrodynamic\\breakup}};

\node[fill=black!20,fill opacity=0,text opacity=0.35] at (axis cs:0.1,0.65) {\small\shortstack{Capillary\\raft}};

\node[fill=teal!50!black!50,fill opacity=0,text opacity=0.35] at (axis cs:10,0.65) {\small\shortstack{Hydrodynamic\\dispersion}};

\end{groupplot}
\end{tikzpicture} 
    \caption{Clustering regimes in the $We_\mathrm{cl}-\phi$ space for open-channel flows, re-defined from \citet{Shin2024DenseInterface}.}
    \label{Fig1}
\end{figure}
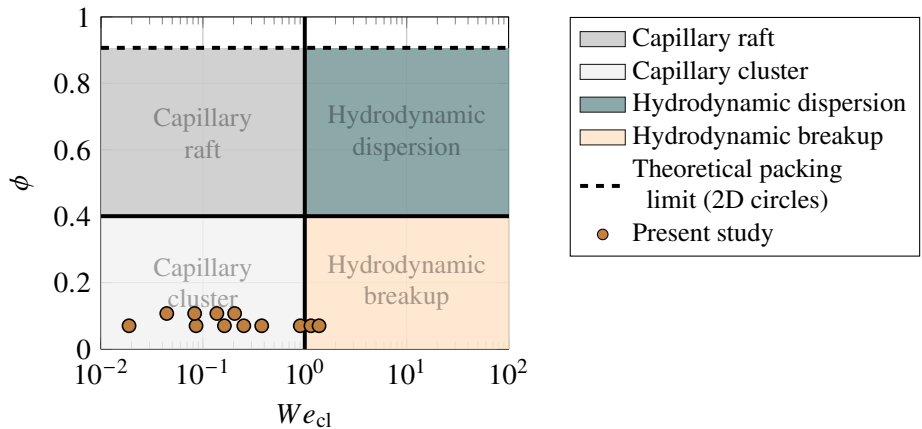

Inspired by this framework, figure~\ref{Fig1} illustrates the clustering regimes for open-channel turbulent flows in the $We_\mathrm{cl}$–$\phi$ parameter space, where $We_\mathrm{cl}$ is the clustering Weber number introduced here to characterise the competition between inertial drag and capillary attraction. The four regimes, defined as capillary cluster, capillary raft, hydrodynamic breakup and hydrodynamic dispersion, capture the dominant mechanisms controlling particle aggregation and breakup. Along the horizontal axis, $We_\mathrm{cl} < 1$ indicates capillary-dominated attraction, whereas $We_\mathrm{cl} > 1$ corresponds to flow forces overcoming capillary attraction. Along the vertical axis, the regimes are separated at $\phi = 0.4$, following \citet{Shin2024DenseInterface}, separating dilute ($\phi < 0.4$) from dense ($\phi > 0.4$) assemblies.

Despite these advances, surface clustering and transport in open-channel flows remain poorly understood. These flows are three-dimensional, with strong turbulence, shear, and channel-scale secondary currents, all of which can influence particle motion and aggregation. In particular, the role of secondary currents in driving lateral particle surface accumulation has not been systematically quantified.

Motivated by this gap, the present study applies the $We_\mathrm{cl}$–$\phi$ framework to open-channel flows. Using controlled flume experiments with two types of buoyant spherical plastic particles differing in size and density, we quantify clustering dynamics and surface distributions through automated image-based analysis. The Weber number $We_\mathrm{cl}$ serves as a measure of the balance between flow-induced disruptive forces and capillary attraction, allowing us to examine how these parameters govern the persistence of particle clusters. Additionally, we demonstrate that channel-scale secondary currents strongly influence lateral accumulation patterns, linking open-channel flow structure to surface transport and providing a mechanistic basis for predicting surface accumulation zones.

\section{Methods}
\label{Methods}
\subsection{\label{Dimensional analysis} Particle clustering framework}

To characterise the clustering behaviour of floating particles in open-channel flows, we adopt a force-based framework that compares inter-particle capillary attraction with the flow-induced disruptive force (hydrodynamic drag). This approach enables assessment of whether particle aggregation can persist under turbulent surface conditions.

\begin{figure}[h]
  \centering
  \includegraphics{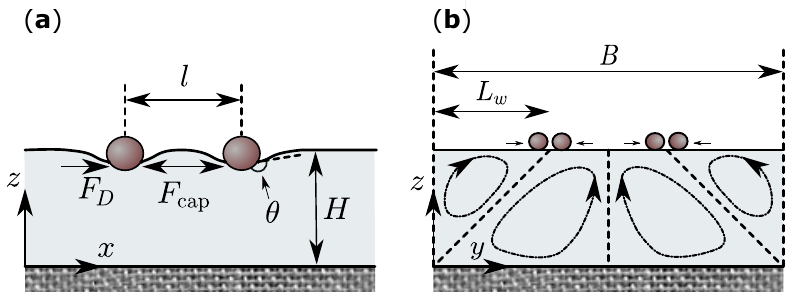}
  \caption{Particle clustering in open-channel flows with water depth $H$; $x, y,z =$ streamwise, spanwise, and vertical coordinate: (\textbf{a}) The interaction between two identical spheres influenced by both capillary ($F_\mathrm{cap}$) and drag forces ($F_D$); here, $l$ is the inter-particle distance and $\theta$ is the contact angle; (\textbf{b}) Schematic of secondary currents induced by the corner between the sidewall and bed. Streamlines represent the mean cross-stream circulation. \(L_w\) denotes the width over which secondary currents develop from the sidewall toward the channel centre.}
  \label{fig:Geometry of a sphere lying at a liquid–gas interface}
\end{figure}

When two particles approach one another, their interfacial deformations overlap and generate lateral capillary attraction (figure~\ref{fig:Geometry of a sphere lying at a liquid–gas interface}a). The resulting horizontal capillary force between two identical particles is written as
\begin{equation}
F_{\text{cap}} = \sigma \, D_p  \, f_\mathrm{cap},
\label{cap force} 
\end{equation}
where $\sigma$ is the surface tension, $D_p$ is the particle diameter, and $f_\mathrm{cap}$ is a dimensionless factor that depends on particle geometry, wetting properties, and inter-particle separation. In the present study, we restrict attention to spherical particles, for which $f_\mathrm{cap}$ is given by
\begin{equation}
f_\mathrm{cap} = \pi  \, Bo^{5/2} \, \Omega^2 \, K_1\left(\frac{l}{L_c}\right), 
\end{equation}
following \citet{vella2005cheerios}. Here, $l$ is the distance between particle centres. The Bond number is defined as \( Bo = (\rho_{w}-\rho_{p})\, g\, D_p^{2} / 4\sigma \),  with $\rho_w$ and $\rho_p$ the densities of water and the particle, and $g$ the gravitational acceleration.
The dimensionless buoyancy-subtracted weight of the particle is 
$\Omega = \frac{1}{3} \left(2\, \frac{\rho_w}{\rho_p} - 1\right) - \frac{1}{2} \cos \theta + \frac{1}{6} \cos^3 \theta,$ where $\theta$ is the contact angle. Finally, $K_1$ is the modified Bessel function of first order, and $L_c = \sqrt{\sigma/(\rho_w g)}$ is the capillary length, which is 2.7 mm for water.

Under flowing conditions, capillary attraction competes with hydrodynamic drag acting on particles at the free surface. This drag arises from the relative velocity between the particle and the surrounding water, that is, the instantaneous slip velocity between the surface flow and the particle 

\begin{equation}
F_D = \frac{1}{2} \rho_w C_D A_{\text{proj},w} \left|u_{\text{fs}} - u_{p}\right| (u_{\text{fs}} - u_{p})
\label{EqDrag}
\end{equation}

where $C_D$ is the drag coefficient, $A_{\text{proj},w}$ is the submerged projected area of the particle, $u_{\mathrm{fs}}$ is the instantaneous fluid velocity at the free surface, and $u_p$ is the instantaneous particle velocity.  Simplifications of (\ref{EqDrag}), as outlined in the Supplementary Material, allow the drag force to be expressed in terms of the bed shear velocity $u_*$, yielding

\begin{equation}
F_{D} = \rho_w \, u_*^2 \, D_p^2 \, f_\mathrm{drag},
\end{equation}
where $f_\mathrm{drag}$ is a dimensionless factor that accounts for particle shape, flow orientation, and inertia. It is given by (Supplemental Material)

\begin{equation}
f_\mathrm{drag} = 0.361 \,  C_D  \,  \left( 1 - \frac{1}{\sqrt{1 + \mathrm{Stk}}}  \right)^2 \frac{A_{\text{proj},w}}{D_p^2} 
\end{equation}
where $\mathrm{Stk}$ denotes the Stokes number, a measure of particle inertia relative to the flow timescale. To quantify the balance between hydrodynamic drag and capillary attraction, we introduce a dimensionless clustering Weber number

\begin{align}
We_\mathrm{cl} &= \frac{F_{\text{D}}}{F_{\text{cap}}} = 
\frac{ \rho_w\, u_*^2\, D_p}{\sigma}  \, \frac{f_\mathrm{drag}}{f_\mathrm{cap}}.
\label{eq.CN}
\end{align}
The numerator of (\ref{eq.CN}) represents flow- and drag-dependent contributions opposing clustering, while the denominator quantifies the particle–interface capillary attraction, with 
$f_\mathrm{drag}$ and $f_\mathrm{cap}$ capturing the effects of particle shape, inertia, and wetting properties, respectively.

Finally, beyond inter-particle force balances, open-channel flows exhibit channel-scale secondary currents arising from the interaction between channel geometry and turbulence. These mean cross-stream circulations act independently of capillary attraction and influence the spatial organisation of floating particles by promoting lateral accumulation in specific surface regions (figure~\ref{fig:Geometry of a sphere lying at a liquid–gas interface}b). The role of secondary currents in shaping surface particle distributions is examined further in section~\ref{sec:implication for real scenarios}.

\subsection{Experimental setup, flow conditions, and cluster detection}
\label{Experimental setup and particle release}

\begin{figure}[h!]
  \centering
  \includegraphics{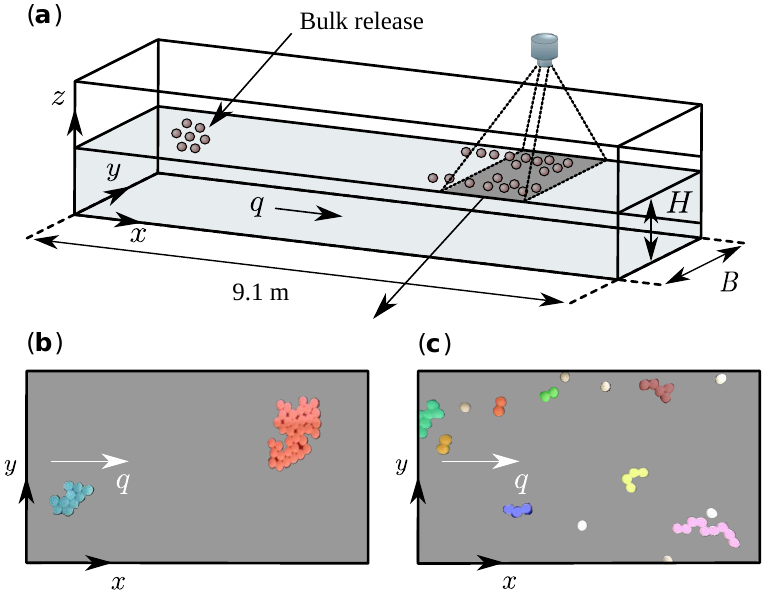}
\caption{Experimental setup and cluster detection: (\textbf{a}) Flume at the Hydraulic Laboratory, UNSW Canberra. Panels (\textbf{b}) and (\textbf{c}) show $D_{30}$ particles under flow conditions $Fr = 0.11$ and $Fr = 0.34$, respectively. 
The clustering percentage $\chi_{\mathrm{cl}}$, as defined in  $\S$ \ref{Results}, is $\chi_{\mathrm{cl}} = 0.97$ (\textbf{b}) and $\chi_{\mathrm{cl}} = 0.85$ (\textbf{c}). Detected clusters are pseudo-coloured for clarity.}
 \label{fig:flume}
 \vspace{-0.5cm}
\end{figure}

The scaling outlined above defines the conditions under which capillary attraction can overcome hydrodynamic drag and sustain clustering. To assess these predictions experimentally, controlled flume experiments were conducted in the Hydraulics Laboratory at The University of New South Wales (Canberra) using a recirculating open-channel flume with internal dimensions of 9.1\,m in length, 0.6\,m in width, and 0.7\,m in height (figure~\ref{fig:flume}\textbf{a}); the sidewalls are made of glass and the bed of acrylic.

Two sets of particles were used to examine the influence of particle properties on clustering behaviour. The first set, \(D_{30}\), comprised 45 polyethylene terephthalate glycol (PETG) spheres ($D_p = 30$ mm, $\rho_p$ = 570\,kg/m$^{3}$), while the second set, \(D_{7}\), consisted of 1000 polypropylene (PP) spheres ($D_p = 7$ mm, $\rho_p$ = 880\,kg/m$^{3}$). For the $D_{30}$ set, eight flow conditions were tested, with the final three conducted at the highest available flow rates. For $D_{7}$, five flow conditions were examined. All flow parameters, including the corresponding Reynolds and Froude numbers, are summarised in Table~\ref{tab:flow_conditions}.

To capture the clustering behaviour, the plastic particles were introduced at the upstream section of the flume using a bulk-release technique. A custom-fabricated circular frame was used to hold the particles in place prior to release, ensuring simultaneous introduction into the flow. Observations were made 6 m downstream of the inlet within a fixed region of interest (figure~\ref{fig:flume}\textbf{a}), where top-view video recordings were acquired using a GoPro camera mounted on an overhead aluminium frame. Each experimental configuration, defined by particle type and flow condition, was repeated three times to ensure repeatability.

\setlength{\heavyrulewidth}{0.8pt} 
\setlength{\lightrulewidth}{0.4pt} 

\captionof{table}{Flow conditions for the experimental sets. The specific discharge $q$ is defined as $q = Q/B$ with $Q$ the flow rate and $B = 0.6$~m the channel width. Reynolds numbers were calculated as $Re = q/\nu$ with $\nu = 1.0\times10^{-6}$~m$^2$/s the kinematic viscosity of water. Froude numbers were calculated as $Fr = q/\sqrt{g H^3}$ with $g = 9.81$~m/s$^2$. The bed shear velocity $u_*$ was computed from the measured mean velocity $U_\mathrm{mean}$ using the log-law proposed by \cite{Swamee1993}, with a roughness height $k_s$ calibrated in detail during a previous PIV measurement campaign; $k_s$  was found to converge to a nearly constant value across previously tested conditions, supporting its use for estimating $u_*$ in this study.}
\begin{center}
\begin{small}
\label{tab:flow_conditions}
\vspace{0.1cm}
\begin{tabular}{c c c c c c c c c}
\toprule
No. & $q$  & $H$  & $B/H$ & $U_\text{mean}$ & $u_*$ & $Re$ & $Fr$ & Tested particles \\
 (-) & (m$^2$/s) & (m) & (-) & (m/s) & (m/s) & (-) & (-) & (-) \\
\midrule
1 & 0.0083 & 0.085 & 7.1 & 0.10 & 0.0058 & $8.3 \times 10^{3}$ & 0.11 & $D_{7}, D_{30}$ \\
2 & 0.023  & 0.107 & 5.6 & 0.22 & 0.011  & $2.3 \times 10^{4}$ & 0.21 & $D_{7}, D_{30}$ \\
3 & 0.043  & 0.140 & 4.3 & 0.31 & 0.015  & $4.3 \times 10^{4}$ & 0.26 & $D_{7}, D_{30}$ \\
4 & 0.077  & 0.190 & 3.2 & 0.41 & 0.019  & $7.7 \times 10^{4}$ & 0.30 & $D_{7}, D_{30}$ \\
5 & 0.11   & 0.220 & 2.7 & 0.50 & 0.023  & $1.1 \times 10^{5}$ & 0.34 & $D_{7}, D_{30}$ \\
6 & 0.057  & 0.095 & 6.3 & 0.60 & 0.029  & $5.7 \times 10^{4}$ & 0.62 & $D_{30}$ \\
7 & 0.082  & 0.115 & 5.2 & 0.71 & 0.033  & $8.2 \times 10^{4}$ & 0.67 & $D_{30}$ \\
8 & 0.12   & 0.150 & 4.0 & 0.80 & 0.037  & $1.2 \times 10^{5}$ & 0.66 & $D_{30}$ \\
\bottomrule
\end{tabular}
\end{small}
\end{center}
\vspace{0.5cm}

Cluster detection was performed using a custom computer-vision pipeline combining object detection, centroid-based tracking, and spatial clustering analysis. Two separate models were developed using the Roboflow platform \citep{roboflow2024} for annotation, data augmentation, and dataset management, and trained using the YOLOv8 architecture \citep{jocher2023ultralytics}.
One model targeted the
\(D_{30}\)  particles, enabling per-particle detection and tracking for frame-by-frame motion analysis, while the second targeted the \(D_{7}\) particles, for which individual detection was less reliable and clusters were identified directly. In both cases, a cluster was defined as a minimum of two visibly contacting particles.

Representative output from the detection pipeline under different flow conditions is shown in figures~~\ref{fig:flume}\textbf{b}, \textbf{c}. These examples demonstrate the effectiveness of our approach in identifying cluster boundaries across particle sizes and velocities, with a detailed discussion provided in $\S$ \ref{Results}.

\section{Results and discussion}
\subsection{Clustering behaviour: when do clusters form?}
\label{Results}

Particle clustering was evident across all tested configurations, with its extent and spatial pattern strongly influenced by both particle and flow characteristics. Representative snapshots from the experiments are presented in figures~\ref{fig:flume}\textbf{b}, \textbf{c}, comparing clustering at two flow velocities for the $D_{30}$  particles. At the lower flow rates, particles formed compact, cohesive clusters (figure \ref{fig:flume}\textbf{b}), whereas at the higher flow rates (figure \ref{fig:flume}\textbf{c}), these clusters became more fragmented and dispersed. Both particle sets exhibited the same qualitative trend, indicating that increasing flow velocity amplifies the disruptive influence of hydrodynamic drag. These observations emphasise the combined effects of flow conditions and particle properties in shaping clustering dynamics at the air–water interface.

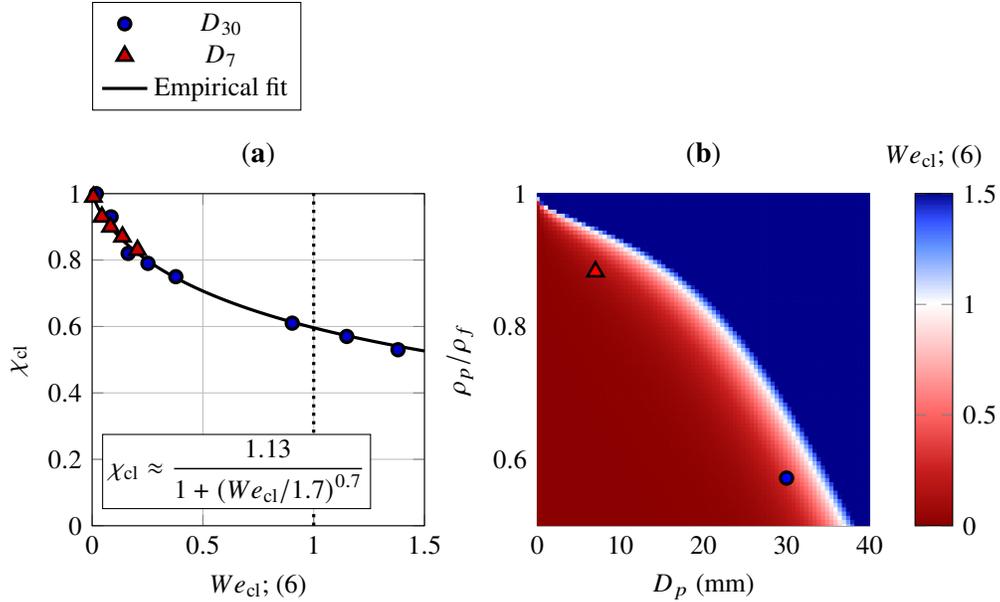
\begin{figure}[h]
\centering
\begin{tikzpicture}

\pgfplotsset{
  every axis/.append style={
    grid=both,
    tick label style={font=\small},
    label style={font=\small},
    width=6 cm,
    height=6 cm,
  },
  /pgf/number format/.cd, 1000 sep={}
}

\pgfplotsset{
  Dthirty/.style={fill=blue, only marks, mark=*,         mark size=2.3pt, line width=1.2pt, draw=black,draw opacity=1},
  Dseven/.style ={fill=red,  only marks, mark=triangle*, mark size=3.5pt, line width=1pt,draw=black,draw opacity=1},
  EmpFit/.style ={black, very thick},
}

\begin{groupplot}[
  group style={group size=2 by 1, horizontal sep=1.5cm, vertical sep=1.5cm},
  legend style={
    at={(0,1.25)},
    anchor=south west,
    legend columns=1,
    draw,
    fill=white,
    font=\small
  }
]

\nextgroupplot[
  xlabel={$We_\mathrm{cl}$; (\ref{eq.CN})},
  ylabel={$\chi_{\mathrm{cl}}$},
  title={(\textbf{a})},
  ymin=0, ymax=1, xmin=0, xmax=1.5,
]

\addplot+[Dthirty] coordinates {
(0.018844, 1.00)
(0.085666, 0.93)
(0.162654, 0.82)
(0.251972, 0.79)
(0.377732, 0.75)
(0.903222, 0.61)
(1.149298, 0.57)
(1.380636, 0.53)
};

(0.022267, 1.00)
(0.097853, 0.93)
(0.185224, 0.82)
(0.288420, 0.79)
(0.430928, 0.75)
(0.969717, 0.61)
(1.237130, 0.57)
(1.496258, 0.53)

(0.009394, 0.99)
(0.066369, 0.93)
(0.124464, 0.90)
(0.205206, 0.87)
(0.306154, 0.83)

\addplot+[Dseven] coordinates {
(0.005969, 0.99)
(0.044208, 0.93)
(0.082962, 0.90)
(0.137086, 0.87)
(0.204829, 0.83)
};

\addplot[black, dotted, line width = 1.2 pt, forget plot] coordinates {(1, 0) (1, 1.05)};


\addplot[EmpFit, domain=0:1.5, samples=300, smooth]
  {1.013/(1+pow(x/1.68,0.69))};

  \node[
  anchor=south west,
  font=\small,
  fill=white,
  draw=black,
  fill opacity=1,
  text opacity=1,
  inner sep=2pt,
]
at (axis description cs:0.03,0.05)
{$\chi_{\mathrm{cl}}\approx \dfrac{1.13}{1+\left(We_\mathrm{cl}/1.7\right)^{0.7}}$};

\addlegendentry{$D_{30}$}
\addlegendentry{$D_{7}$}
\addlegendentry{Empirical fit}

%

\nextgroupplot[
 title={(\textbf{b})},
    xlabel={$D_p$ (mm)},
    ylabel={$\rho_p/\rho_f$},
    xmin=0, xmax=40,
    ymin=0.5, ymax=1.0,
    colormap name=mygray, 
    point meta min=0,
    point meta max=1.5,
    colorbar,colorbar style={
    title=$We_\mathrm{cl}$; (\ref{eq.CN}),  
    title style={font=\small},
    tick label style={font=\small}
 },
]

\addplot[
  matrix plot*,
  mesh/cols=80,            
  point meta=explicit,
] table[
  col sep=space,
  header=true,
  x=D_mm,
  y=rho_ratio,
  meta=We_cl,
] {Ca_map_pgf.dat};

\addplot [Dthirty]
  coordinates {(30,0.5717)}; %

\addplot[Dseven]
  coordinates {(7,0.8826)};  %

\end{groupplot}

\end{tikzpicture}

\caption{Clustering behaviour of buoyant particles: (\textbf{a}) Weber number $We_\mathrm{cl}$ as a function of the fraction of clustered
particles $\chi_{\mathrm{cl}}$ for $D_{30}$ and $D_{7}$ particles; the solid curve is an empirical fit to the combined dataset; (\textbf{b}) Weber number $We_\mathrm{cl}$ as a function of particle diameter $D_p$ and density ratio $\rho_p/\rho_f$ for $q= 0.11$ m$^2/s$ and $Fr = 0.34$; the contact angle is taken as $\theta = 95^\circ$.}
\label{fig:cluster-colapse}
\end{figure}

To quantify the extent of clustering across experimental conditions, we define the fraction of clustered particles, $\chi_{\mathrm{cl}}$

\begin{equation}
\chi_{\mathrm{cl}} = \left( \frac{N_{p,\mathrm{total}} - N_{p,\mathrm{free}}}{N_{p,\mathrm{total}}} \right),
\label{eq:percent_clustered}
\end{equation}
where \( N_{p,\mathrm{total}} \) is the total number of particles in the experimental set, and \( N_{p,\mathrm{free}} \) is the number of particles that remain isolated (not belonging to any cluster).  Figure~\ref{fig:cluster-colapse}\textbf{a} compares the experimentally measured $\chi_{\mathrm{cl}}$ with the Weber number $We_\mathrm{cl}$, evaluated using (\ref{eq.CN}),
showing a clear monotonic decrease of $\chi_{\mathrm{cl}}$ with increasing $We_\mathrm{cl}$. For the evaluation of (\ref{eq.CN}), the inter-particle separation, surface tension and the drag coefficient are taken as $l = D_p$, $\sigma = 0.072$ N/m, and  $C_D \approx 0.28$ \citep{kamoliddinov2021hydrodynamic}, respectively, while the projected area $A_{\text{proj},w}$ and the contact angle are evaluated on the basis of particle submergence, which is determined from image-based analysis of side-view images of the particles at the air–water interface under hydrostatic conditions. The contact angles are obtained from an inverse vertical force balance; see, e.g., \cite{lee2018static}. They yield $\theta_{7} \approx 86^\circ$ and $\theta_{30} \approx 95^\circ$.

For $We_\mathrm{cl} \lesssim 0.1$, particles remain strongly clustered ($\chi_{\mathrm{cl}} \approx 1$), whereas a rapid decline occurs as $We_\mathrm{cl}$ approaches unity, marking the transition to drag-dominated behaviour. For $We_\mathrm{cl} > 1$, clusters become weak and transient, with fewer than 60 \% of particles remaining aggregated. Data from both $D_{30}$ and $D_{7}$ particles collapse onto a single trend, demonstrating that $We_\mathrm{cl}$ captures the dominant physics despite differences in particle size and properties. This collapse confirms that the clustering dynamics can be described by a single dimensionless force balance, with $We_\mathrm{cl} \approx 1$ representing the threshold at which drag and capillary forces become comparable. The agreement of both particle sets with this trend provides strong experimental support for the proposed scaling framework.

While figure~\ref{fig:cluster-colapse}\textbf{a} demonstrates the usefulness of the Weber number $We_\mathrm{cl}$ as a single control parameter, figure~\ref{fig:cluster-colapse}\textbf{b} provides insight into the distinct roles of particle size and density in controlling clustering behaviour. In this panel, $We_\mathrm{cl}$ is mapped as a function of particle diameter $D_p$ and density ratio $\rho_p/\rho_f$ for a fixed flow condition ($q=0.11$ m$^2$/s and $Fr = 0.34$). Increasing $D_p$ shifts the system towards higher values of $We_\mathrm{cl}$, as larger particles experience stronger hydrodynamic drag due to their increased projected area $A_\mathrm{proj}$. 
Particle density $\rho_p$
also influences $We_\mathrm{cl}$
through multiple mechanisms. Heavier particles have larger Stokes numbers ($\mathrm{Stk}$) that increase the drag force ($F_D$) and destabilise clusters, while lighter particles experience stronger capillary attraction through larger Bond numbers ($Bo$) and higher dimensionless weights ($\Omega$), promoting clustering.
Taken together, particle size and density act in a complementary manner: size controls how rapidly drag becomes dominant, while density sets whether capillary attraction is sufficient to sustain clusters. This interplay explains the distinct clustering responses observed experimentally for the $D_{30}$ and $D_{7}$ particles and reinforces the interpretation of $We_\mathrm{cl}$ as a physically meaningful measure of the competing capillary and drag forces at the interface.

\subsection{Surface distribution: where do clusters persist?}
\label{sec:implication for real scenarios}

Although the metric $We_\mathrm{cl}$ quantifies the clustering behaviour of floating particles, it does not capture the mechanisms that govern their spatial distribution across the free surface. Previous studies have examined clustering at the local scale by correlating particle patterns with the instantaneous surface velocity divergence \citep{lovecchio2013time}. In contrast, the present experiments focus on large-scale lateral positioning driven by secondary currents of Prandtl’s second kind, which arise from interactions between wall and bed boundary layers (figure~\ref{fig:Geometry of a sphere lying at a liquid–gas interface}\textbf{b}) \citep{prandtl1925uber, einstein1956viscous,perkins1970formation}. The near-surface cross-stream motion of these circulations can redistribute floating particles across the channel width, leading to preferential accumulation at specific spanwise positions depending on the channel aspect ratio $B/H$. On this basis, three general regimes of surface particle distribution can be identified, in accordance with the seminal classification for smooth open-channel flows proposed by \cite{nezu1993turbulence}
\begin{itemize}
   \item \textbf{Narrow channels} ($B/H \leq 2$)\\
    Two opposing secondary-current cells merge near the channel centre, causing particles to travel along and accumulate at the centreline.

    \item \textbf{Intermediate aspect ratios} ($2 \leq B/H \lesssim 6$) \\
    Two accumulation zones form near the sidewalls, caused by the outward-flowing surface branches of corner-induced secondary currents, which guide particles to specific lateral positions (figure \ref{fig:Geometry of a sphere lying at a liquid–gas interface}\textbf{b}).

    \item \textbf{Wide channels} ($B/H \gtrsim  6$)\\
    Particle distributions are approximately uniform across the channel width, indicating weak mean cross-stream transport at the free surface. In this regime, secondary currents are confined near the walls or are too weak to impose a persistent surface convergence pattern across the full width.
\end{itemize}

It should be noted that, in wide channels, multiple secondary-current cells can develop across the channel width, particularly in the presence of bed roughness variations \citep{zampiron2020secondary, rodriguez2008laboratory}, which can modify the organisation of the currents and, consequently, the resulting surface particle distribution.

\begin{figure}[h!]
\centering
\begin{tikzpicture}
\begin{groupplot}[
    group style={
        group size=2 by 1,  
        horizontal sep=1.1 cm, 
    },
    tick label style={font=\normalsize},
    label style={font=\normalsize},
]

\nextgroupplot[
title={(\textbf{a})},
    axis lines=none,
    ticks=none,
    xlabel={}, ylabel={},
    width=9cm,   
    height=6.2cm, 
    clip mode=individual,  
]

\node[inner sep=0pt, anchor=south west] at (rel axis cs:0,-0.02)
{\includegraphics[width=7.1 cm,height=4.7cm]{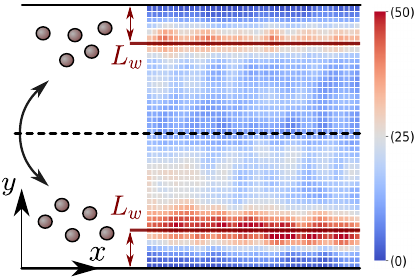}};

\node[inner sep=0pt, anchor=south west] at (rel axis cs:0.81,1.05)
{Count};

\addplot [draw=none] coordinates {(0,0)};  

\nextgroupplot[
    title={(\textbf{b})},
    xlabel={$ (L_w)_\text{meas}/ H$},
    ylabel={$ (L_w)_\text{calc}/ H$; (\ref{eq:yang-slope})},
    xmin=0, xmax=1.5, ymin=0, ymax=1.5,
    grid=both,
    width=6. cm,   
    height=6.cm,  
    scaled ticks=false,
    legend style={
        font=\small,
        draw=black,
        at={(1,0)},
        anchor=south east
    },     legend image post style={scale=1},
]

\addplot[black, thick, mark=square*, only marks, mark size = 2] coordinates {({0.18/0.22},{0.22/0.22})};
\addlegendentry{$B/H=2.7$}

\addplot[black, thick, mark=triangle*, only marks,mark size = 2.5] coordinates {({0.167/0.19},{0.2/0.19})};
\addlegendentry{$B/H=3.2$}

\addplot[black, thick, mark=*, only marks,mark size = 2] coordinates {({0.146/0.14},{0.159/0.14})};
\addlegendentry{$B/H=4.3$}

\addplot[black, thick, mark=diamond*, only marks,mark size = 2.5] coordinates {({0.133/0.107},{0.125/0.107})};
\addlegendentry{$B/H=5.6$}

\addplot[black, dashed, thick, domain=0:1.5] {x};

\end{groupplot}

\end{tikzpicture}
\caption{Surface distributions of buoyant plastics:
(\textbf{a}) Ensemble-averaged surface count of buoyant particles ($D_7$) for an intermediate channel aspect ratio $B/H = 4.3$; colours indicate the number of particle detections per surface grid cell, with red regions corresponding to persistent accumulation zones;  
(\textbf{b}) Comparison between calculated and measured $L_w$, normalized by the respective water depth $H$.}
\label{fig:expertimental data}
\end{figure}
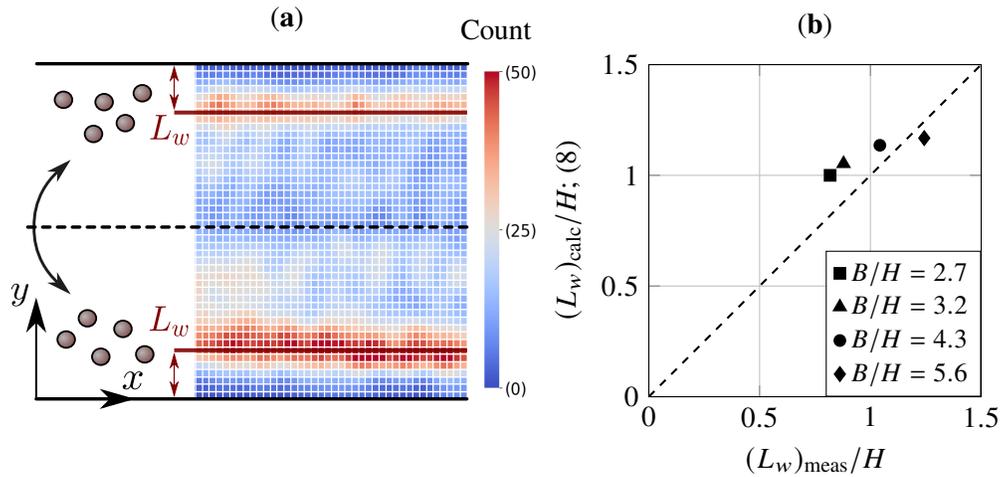

To assess how secondary currents influence surface transport, we examine the lateral distribution of the $D_7$ particles across the flow configurations in Table~\ref{tab:flow_conditions} for $B/H < 6$. The $D_{30}$ particles were not included in this analysis, as only 45 spheres were available, which was insufficient for statistically meaningful lateral distributions. As an illustrative example, figure~\ref{fig:expertimental data}\textbf{a} shows an ensemble-averaged surface count for an intermediate aspect-ratio case ($B/H =4.3$), obtained by aggregating particle detections at the free surface, with the red bands indicating the observed accumulation zones.

The lateral extent of the near-surface influence of wall-driven secondary currents can be estimated using an analytical solution derived by \citet{yang1997mechanism}, which is valid for smooth  rectangular channels with intermediate aspect ratios ($2 \leq B/H \lesssim 6$).  In this formulation, $L_w$ represents the surface width over which the secondary currents develop from the sidewall toward the channel centre

\begin{equation}
    \left( \frac{L_w}{H} \right)^3 + \left( \frac{2H}{B} \right) \frac{L_w}{H} - 2 = 0, \qquad \text{for } B/H \geq 2,
    \label{eq:yang-slope}
\end{equation}
and \(L_w/H\) denotes the slope of the interface separating the wall- and bed-driven shear layers. Figure~\ref{fig:expertimental data}\textbf{b} compares the theoretically predicted $L_w$ values with our experimental measurements. For intermediate aspect-ratio cases, the observed accumulation bands closely match the predicted lateral extent, indicating that the analytical solution provides a reasonable estimate of where surface particles preferentially accumulate in smooth rectangular channels. This finding demonstrates that the structure of secondary currents can be used to predict lateral particle distributions, linking channel geometry directly to surface transport patterns and offering a mechanistic basis for anticipating accumulation zones in natural and engineered waterways.

\subsection{Limitations and future outlook}
While the present study provides insight into particle clustering and lateral transport, several limitations should be noted. The experiments were conducted in a smooth, straight, rectangular channel, and the analytical predictions assume fully developed secondary currents; natural channels with rough beds, bends, or obstacles may produce different accumulation patterns. The analysis also focuses on mean flow effects, neglecting transient turbulent fluctuations that could intermittently influence clustering and lateral transport. Additionally, only spherical particles were investigated, while partial submergence and surface contamination were not systematically varied, which may affect capillary attraction and drag under natural conditions.

Future work should extend the present study to a broader range of experimental conditions. This includes introducing variations in bed roughness, testing different particle shapes and combinations thereof, and exploring additional flow configurations. Investigating both narrower ($B/H \leq 2$) and wider channels ($B/H \gtrsim 6$) would further clarify the generality of the findings across different flow regimes.


\section{Conclusion}
\label{Conclusion}

This study investigated the clustering and spatial organisation of buoyant plastic particles at a free surface under open-channel turbulence through a force-based theoretical framework and controlled flume experiments. We introduced a dimensionless clustering Weber number ($We_\mathrm{cl}$) for open channel flow, which compares hydrodynamic drag with capillary attraction, and showed that it provides a consistent description of clustering behaviour across particle types and flow conditions.

Beyond the extent of clustering, our analysis revealed that the lateral positioning of clusters is governed by secondary current intrinsic to open-channel flows. Ensemble-averaged surface counts across varying aspect ratios showed that, for intermediate aspect ratios ($2 \lesssim B/H \lesssim 6$), the observed accumulation bands closely match the analytically predicted lateral extent of secondary currents, confirming that buoyant particles tend to follow the surface streamlines of these secondary circulations.

Together, these findings establish a framework in which the dimensionless Weber number $We_\mathrm{cl}$ determines when aggregation occurs, while the secondary circulation structure set by the channel geometry dictates where clusters persist. This dual perspective connects small-scale interfacial physics with large-scale flow organisation, offering predictive capability for floating-particle accumulation in waterways.

\medskip
\noindent\textbf{Data Availability Statement.}
Data, models, and code supporting this study are available from the corresponding author upon reasonable request.

\medskip
\noindent\textbf{Acknowledgements.}
The authors thank Felipe Condo-Colcha for providing his PIV measurements. The authors also acknowledge the use of the AI language model ChatGPT (OpenAI) for minor editorial assistance; all interpretations and conclusions are solely those of the authors.

\medskip
\noindent\textbf{Declaration of Interest.}
The authors declare no conflict of interest.

\bibliographystyle{jfm}
\bibliography{bibliography}

\section*{Supplementary material}
\label{Supplementary material}
\setcounter{equation}{0}
\renewcommand{\theequation}{S\arabic{equation}}

This supplementary material outlines the derivation of the drag force acting on a floating particle in open-channel flow, as used in the clustering number formulation of the manuscript.

The force acting on a particle in a flow includes drag contributions as well as unsteady terms, such as added-mass and history forces \citep{crowe2011multiphase}. In the present analysis, only the drag contribution is retained, while unsteady forces are neglected. Under this assumption, the force acting on the particle reduces to a quasi steady-state drag determined by the instantaneous velocity between the particle and the free-surface flow

\begin{equation}
F_D = \frac{1}{2} \rho_w C_D A_{\text{proj},w} \left|u_{\text{fs}} - u_{p}\right| (u_{\text{fs}} - u_{p}),
\label{A1}
\end{equation}
where $C_D$ is the drag coefficient, $A_{\text{proj},w}$ is the submerged projected area, $u_{\text{fs}}$ is the instantaneous fluid velocity at the free surface, and $u_p$ is the instantaneous particle velocity.

We decompose the fluid and particle velocities at the particle location using a Reynolds decomposition, $u_\text{fs} = \overline{u}_\text{fs} + u_\text{fs}^\prime$ and $u_\text{p} = \overline{u}_\text{p} + u_\text{p}^\prime$. Here, we assume statistically steady conditions in which any mean slip velocity has relaxed, so that $\overline{u}_\text{fs} = \overline{u}_p$, and relative motion arises solely from velocity fluctuations

\begin{equation}
F_{D} = \frac{1}{2} \rho_w C_D A_{\text{proj},w} \left| u_\text{fs}^\prime - u_p^\prime\right| ( u_\text{fs}^\prime - u_p^\prime).
\label{S2}
\end{equation}

To relate particle and free-surface velocity fluctuations, we use a first-order response model in which the particle velocity relaxes toward the carrier-flow velocity over the Stokes response time \citep{crowe2011multiphase}. This behaviour can be represented as

\begin{equation}
u_p' \approx \frac{u_{\mathrm{fs}}'}{\sqrt{1+\mathrm{Stk}}},
\label{S3}
\end{equation}
where the Stokes number is defined as $\mathrm{Stk} = \tau_p / \tau_{\mathrm{fs}}$, with $\tau_p$ the particle relaxation time

\begin{equation}
\tau_p = \frac{\rho_p D_p^2}{18 \mu},
\end{equation}
computed from Stokes drag to provide a characteristic particle response timescale. Despite the instantaneous drag being quadratic, this approach correctly captures particle response and is standard in turbulence–particle models. The free-surface turbulent timescale is

\begin{equation}
\tau_\text{fs} = \frac{\mathcal{L}_\text{fs}}{u_\text{fs}'} = \frac{\alpha \, H}{u_\text{fs}'},
\end{equation}
where $\rho_p$ and $D_p$ denote the particle density and diameter, respectively, $\mu$ is the dynamic viscosity, and the free-surface length scale is taken as $\mathcal{L}_{\mathrm{fs}} = \alpha H$, where $\alpha$ is an empirical factor that relates $\mathcal{L}_{\mathrm{fs}}$ to the flow depth $H$. Experimental and numerical studies of open-channel turbulence indicate that $\alpha \approx 0.4$ \citep{nezu1993turbulence,bauer2025freesurface}, yielding

\begin{equation}
\mathrm{Stk} = \frac{\rho_p D_p^2 \, u_\text{fs}'}{18 \mu \, \alpha \, H}.
\end{equation}

The Stokes number provides physical insight into the fluid–particle interaction: for $\mathrm{Stk} \ll 1$, the particle follows the flow closely ($u_p' \approx u_\text{fs}'$), whereas for $\mathrm{Stk} \gg 1$, the particle responds weakly to flow fluctuations ($u_p' \ll u_\text{fs}'$). Substituting (\ref{S3}) into (\ref{S2}) and approximating the instantaneous fluctuation magnitude by its root-mean-square value, $u_\text{fs}^{\prime 2} \approx u_\text{fs,rms}^{\prime 2}$, yields

\begin{equation}
F_{D} = \frac{1}{2} \rho_w C_D A_{\text{proj},w} \, u^{\prime 2}_\text{fs,rms} \left( 1 - \frac{1}{\sqrt{1 + \mathrm{Stk}}} \right)^2,
\label{S7}
\end{equation}
where this approximation captures the dominant contribution to the mean drag while neglecting higher-order moments of the velocity fluctuations. Using a semi-empirical relationship for root-mean-square velocity fluctuations in open-channel flows proposed by \cite{nezu1993turbulence} and evaluating it at the free surface ($z = H$), we obtain

\begin{equation}
u_\text{fs,rms}^{\prime} = 2.30 \, u_* \, \exp\left(-\frac{z}{H}\right) = 2.30 \, u_* \, \exp(-1) \approx 0.85 \, u_*,
\label{S8}
\end{equation}
where $u_*$ is the bed shear velocity. Finally, combining (\ref{S7}) and (\ref{S8}) gives an expression for the drag force acting on a floating particle

\begin{equation}
F_{D} = 0.361 \, \rho_w C_D A_{\text{proj},w} \, u_*^2 \, \left( 1 - \frac{1}{\sqrt{1 + \mathrm{Stk}}} \right)^2.
\end{equation}

\end{document}